# Complementing Carbon Credits from Forest-Related Activities
## with Biodiversity Insurance and Resilience Value
### *PREPRINT*
Word count main text: 7899
Word count abstract: 133 words

Hanna Fiegenbaum[a]

[a]*Leipzig University, Institute for Medical Informatics, Statistics and Epidemiology, Leipzig, Germany, hanna.fiegenbaum@uni-leipzig.de, hanna.fiegenbaum@gmail.com*

**Abstract:** Carbon credits are a key component of most national and organizational climate strategies. Financing and delivering carbon credits from forest-related activities faces multiple risks at the project and asset levels. Financial mechanisms are employed to mitigate risks for investors and project developers, complemented by non-financial measures such as environmental and social safeguards and physical risk mitigation. Despite these efforts, academic research highlights that safeguards and climate risk mitigation measures are not efficiently implemented in some carbon projects and that specification of environmental safeguards remains underdeveloped. Further, environmental and social risk mitigation capacities may not be integrated into financial mechanisms. This text examines how ecosystem capacities can be leveraged and valued for mitigation of and adaptation to physical risks by complementing carbon credits with biodiversity insurance and resilience value.

## Introduction
### Nature-based Credits as a Financing Instrument to attain Nature and Climate Targets
The Paris Agreement and the Kunming-Montreal Global Biodiversity Framework set environmental targets to halt climate change and reverse biodiversity loss. Both agreements rely on the public and private sector's contributions to attain climate and nature targets, not only through implementation of measures to avoid and reduce emissions and further degradation of the state of nature, but also through the provision of public and private capital to finance nature-based solutions with goals such as increasing carbon sequestration and the provision of other ecosystem services, nature restoration and to reverse biodiversity loss (Deutz et al., 2020). Nature-based credits are a class of market-based financing instruments used to fund carbon avoidance or removal through nature-based carbon credits (e.g. Haya et al., 2024; West et al., 2020), or in the case of biodiversity credits, to fund biodiversity conservation and net gains (e.g. Wunder et al., 2024; zu Ermgassen et al., 2019, 2020, 2023).

There exist voluntary and compliance markets for nature-based carbon removal and avoidance projects. An example for carbon projects involving forestry in a compliance market is the California cap-and-trade program (e.g. Badgley et al., 2022; Haya et al., 2020, 2023). Credits in compliance markets are known as allowances or offset credits, and they are used by entities to meet legally mandated emission reduction targets. Credits in voluntary carbon markets are purchased by countries, companies and organizations to be used for compliance under national or regional carbon markets and targets or in the case of companies, as part of voluntary carbon offsetting or enhancing their corporate carbon balance within their CSR commitment to achieve carbon neutrality or climate positive targets. Institutions and mechanisms to enable financing of nature-based climate solutions to either avoid deforestation or enable afforestation and reforestation through carbon credits have been in development for decades. This includes the Clean Development Mechanism followed by the Sustainable Development Mechanism included in the Paris Agreement, the REDD and REDD+ initiatives to reduce deforestation and promote sustainable development, numerous national or regional off-setting markets and the Voluntary Carbon Market overseen by bodies such as the Integrity Council for the Carbon Market and the Voluntary Carbon Markets Integrity Initiative.

Carbon credits focus on financing the avoidance and removal of carbon emissions, assessed in a metric of tons of carbon dioxide equivalent. They operate on an equivalence principle making emissions, their avoidance, reduction or removal from different sources commensurable (Carton et al., 2021; IPCC sixth assessment, 2022; Smith et al., 2017). This allows companies or countries to integrate carbon credits from different sources into their carbon balances and accounting while mobilizing private and public capital to fund the implementation of nature-based climate solutions and other climate mitigation and adaptation measures. Although research has revealed a temporary decline of carbon uptake by terrestrial ecosystems during 2023 (Ke et al., 2024), their functioning as natural carbon sinks remains critical for moderating climate change (Pan et al., 2024). Specifically the conservation, restoration and sustainable management of forests of high diversity offers valuable potential to contribute to meeting global climate and biodiversity targets ( Liang et al., 2016; Mo et al., 2023; Weiskopf et al., 2024) and will rely on outcome-based financing through carbon credits as one crucial financing vehicle (Swinfield et al., 2024; Wells et al., 2023). Academic literature has researched market design and mechanisms of carbon markets since the emergence of their institutions, both in compliance and voluntary markets (Haya et al., 2023; Miltenberger et al., 2021; Pan et al., 2022; Sills et al., 2017; West et al., 2020, 2024). While institutional, regulatory and geographical contexts of carbon projects and programs involving forestry differ (Börner et al., 2020), voluntary and compliance markets for carbon credits generated from forest projects share not only the same commodification mechanism and basic requirements in the carbon credit supply chain leading to the release of credits. They also face similar challenges of measuring, mitigating or commodifying the risks and uncertainties inherent to processes leading to their issuance.

**Financing pathway and quality requirements**
The pathway of financing nature-based climate solutions through credits requires investors to finance projects such as afforestation and reforestation upfront, while credits are ideally issued only if impacts have been reliably produced, measured, reported and verified. The need for prior investment into nature-based project implementation necessitates forecasts of future project delivery. Credit suppliers are encouraged to retreat to conservative estimates of impact delivery whenever uncertainties remain (Core Carbon Principles; European Carbon Removal Framework; Haya et al., 2023).
A key requirement in impact evaluation of project outcomes is their additionality, that the impact created by a project wouldn't have occurred without the project's implementation. Assessing additionality requires establishing a counterfactual baseline against which additionality of the project outcomes is assessed, including control of biases (Badgley et al., 2022; Swinfield and Balmford, 2023; Swinfield et al., 2024; West et al., 2020).This additional impact is required to be adjusted according to the potential leakage that the project might have caused (Aukland et al., 2003; Schwartzman et al., 2021). Leakage occurs when, for instance, protecting or restoring forests in one area leads to increased deforestation or emissions elsewhere, negating or reducing the benefits of the project. Assessing the additionality, leakage, and durability of carbon storage relative to a counterfactual baseline along with maintaining rigorous carbon accounting on the supply as well as on the demand side, are key quality requirements for carbon credits and the protocols and methodologies used in their issuance (Kaplan et al., 2023; Pietracci et al., 2023; Swinfield and Balmford, 2023). Projects are implemented and monitored over time to ensure that they meet the necessary standards and actually deliver the promised impact. Once the impacts or outcomes are verified by an independent third party, the project is certified, and credits are issued. Investors or companies that finance the project can sell the issued credits on the market or use them for their own carbon accounting.

**Risks and risk mitigation measures**

There exist risks that carbon sequestration is reversed after project implementation, due to natural or human factors, which undermines the long-term effectiveness of carbon credits. A tension exists in nature-based carbon removal within the equivalence principle through the fact that carbon dioxide emissions from fossil fuels persist for hundreds to thousands of years (Archer et al., 2009; Joos et al., 2013), while carbon stored in natural sinks is inherently less permanent and is subject to significant risks, which are expected to grow in a changing climate (Anderegg et al., 2013; 2020; 2022; Barbero et al., 2015). A common measure to address the risk of impermanence is the creation of buffer pools of carbon credits which are used in voluntary markets as well as in compliance markets such as in the forest offset projects in the California Cap-and-Trade Program (Badgley et al., 2022). A buffer pool sets aside a portion of carbon credits as a reserve to cover potential carbon losses due to unforeseen events, such as wildfires or disease, ensuring the overall integrity and reliability of the program's climate impact. Additional risks for investors arise at the project level from potential non-delivery of credits, overestimation of delivery, contestable claims and long time lags between investment and return. Sources of risks that have been identified to successful project delivery include nature-related risks from diseases, pests, tree mortality or biodiversity loss (Anderson-Texeira et al., 2022; Mori et al., 2021), climate-related risks such as wildfires, droughts, floods or storms and the uncertain forest responses to those (Anderegg et al., 2022, 2020; Anderson-Texeira et al, 2022, Galik and Johnson, 2009), risks from non-compliance such as contract breach, logging or clear-cutting (Fujii et al., 2024; Gifford, 2020), or from unclear liabilities and property rights and jurisdictional weakness such as lack of enforcement (Chan et al., 2023; MacKenzie et al., 2012) or regulatory changes (Dutschke et al., 2004).

Several mechanisms are thus in place to assess and manage financial as well as environmental and social risks for investment in nature-based carbon removal projects. Conducting due diligence is a critical component of understanding not only financial, but also social and environmental risks of a project. Due diligence ensures that projects meet standards and align with sustainable development goals and its outcome is factored into the decision-making process for proceeding with a transaction. Further, staged investments imply that funding is released incrementally based on project milestones. Mechanisms such as certification and independent verification, guaranteed offsets, insurance and hedging can be used to address financial risk (Chan et al., 2023; Rau et al., 2024b; Tarnoczi, 2017). They seek to reduce financial risk of an investment through risk transfer and distribution. Further suggestions for mitigating risks from the academic literature include robust baselining (Swinfield et al., 2024; West et al., 2020), retreating to conservative estimates of project delivery (Haya et al., 2023; Swinfield et al., 2024), as well as using optimization of release schedules for credits that limit the risk of credit reversal (Rau et al., 2024a) and the risk failing to generate credits (Rau et al., 2024c).

**Limitations of financial risk mitigation**

While financial risk mitigation strategies address the potential financial loss of investors from possible credit non-delivery or reversal, they don't necessarily address the physical and social risks, the potential environmental and social damage from project failure. This is due to the limited substitutability of the different capitals, e.g. financial and natural capital. The distinction between financial, social and environmental risks is captured in the concept of double materiality. Compensating financially for the damage loss might not help restore ecosystems from irreversible damage. Further, by financially insuring against the loss, risk is distributed in society rather than reduced or its potential material impacts minimized.

To address social and environmental risks to a project, most carbon credit programs and standards have therefore implemented project requirements to reduce and mitigate physical and social risks to projects. They establish environmental and social safeguards (Wissner & Schneider, 2022) and include project-level risk mitigation in methodologies (e.g. Gold Standard, 2013). The social and environmental safeguard approach is sometimes described as a "do-no-harm" policy. Social and environmental safeguards are policies, procedures, or measures put in place to prevent or minimize negative social and environmental impacts associated with projects, programs, or investments. These safeguards are essential to ensure that economic development or resource management efforts are conducted responsibly and sustainably and promote positive outcomes on a variety of social, economic and environmental dimensions and are therefore included in sustainable finance frameworks such as the EU Taxonomy. Social safeguard policies have been developed with respect to the inclusion and participation of indigenous people and local communities in carbon project design and development (see Bonfante et al., 2010 with a focus on safeguards in REDD+ projects). This effort is supported by the academic literature which highlights that project success is conditional on social-ecological embeddedness and enhanced local participation, along with community governance and oversight at the project level (Griffiths et al., 2019, 2020; Löfqvist et al., 2023; zu Ermgassen & Löfqvist, 2024). Additionally, it aligns with equity and justice principles that necessitate the participation and leadership of indigenous people and local communities in project planning and implementation to ensure positive local impact of projects in livelihood, equity and justice (Herr et al., 2019; McDermott et al., 2013).

**Shortcomings in implementing safeguards and adaptation measures**

However, academic literature still reports on shortcomings in implementing social safeguarding measures as well as in addressing climate risks to permanence in some projects (Ashraf & Karaki, 2024; Haya et al., 2024). Further, environmental safeguards are rarely discussed. Wissner and Schneider (2022) therefore recommend developing specified safeguards addressing risks to specific environmental components such as biodiversity. This recommendation aligns with the academic literature which suggests integrating forest management activities such as mechanical thinning or fuel reduction treatments as a preventive measure to reduce risk from forest fires into adaptation and risk mitigation management on the project level (Daigneault et al., 2010; Herbert et al., 2022; Hurteau et al., 2008). In a similar line of thinking, studies propose transitioning to an adaptation and resilience management approach in forest-based carbon project development to help build robustness and flexibility into systems to handle unexpected challenges and to promote long-term sustainability of projects (Wassenius and Crona, 2022; Wells et al., 2023). So, while the use of financial mechanisms such as insurance can manage and mitigate financial risks, social-ecological resilient management can contribute to incrementing the general resilience of an ecosystem in the face of multiple risks from known to yet unknown disturbances which latter cannot be insured financially. Different risk mitigation measures are applicable to manage, reduce or adapt to risks of different types - financial, climate-related, environmental and social - and can be integrated into a comprehensive strategy combining financial and capacity-related measures of risk mitigation.

**Co-benefits and coupled ecosystem components**

One possible pathway for enhanced integration of adaptation and resilience measures is through their inclusion into project co-benefits. Carbon projects often seek to deliver co-benefits such as the enhancement of local biodiversity, improvement of livelihoods and equity and justice (Lam et al., 2024; Swinfield & Balmford, 2023). In the voluntary carbon market, co-benefits are usually not priced as separate value items adding to the price but can rather increase the perceived value of the credit itself, allowing projects with co-benefits to set a higher base price (Lou et al., 2022, 2023). In compliance or regulated carbon markets, co-benefits are not priced in. This also implies that ecological components other than carbon storage and sequestration remain un- or undervalued. Moreover, on the level of the financing mechanism, carbon and co-benefit components remain essentially decoupled from one another. In contrast, the scientific literature stresses the interconnectedness of ecological components. Diaz et al. (2009) point to biodiversity in carbon projects as not just a decoupled "side benefit", but refer to its stabilizing and productivity-enhancing role in forest ecosystems. Several studies demonstrate that carbon storage potential of forests as well as their general productivity increases proportionate to their tree and plant species diversity (e.g. Liang et al., 2016; Liu et al., 2018; Mori et al., 2021; Weiskopf et al., 2024). Other studies stress the stabilizing effect of tree diversity on the forest carbon cycle and carbon storage (Qiao et al., 2023; Silva Pedro et al., 2015; Zhang et al., 2024;). Sullivan et al. (2017) warn that the distribution of carbon and biodiversity in carbon management of tropical forests has to be considered carefully to avoid trade-offs. While the scientific representation of domain knowledge around carbon removal aims for comprehensiveness, accuracy, consistency, and methodological rigor in examining ecological components and their relationships, its commodification depends on abstraction.

**Challenges from differing domain knowledge representations**

Domain knowledge of nature-based climate solutions is hence represented with varying perspectives and objectives on different decision-making levels and in different fields. After all, the knowledge used in decision-making and activities of land and forest management to manage or steward land and forest for carbon credit delivery, is what is eventually decisive on how biodiversity, carbon and other ecosystem components are managed considering the manifold ways in which they are coupled. Still, research discusses how differing domain knowledge representations as well as their conflicts influence land management decisions by being implicit to carbon accounting, market and financing mechanisms (Buongiorno et al., 2015, Diaz et al., 2018), to policy and regulations (von Hedemann et al., 2021), to carbon program design (Badgley et al., 2022; Gren & Aklilu, 2016; Haya et al., 2020), or protocol design (Marino and Bautista, 2022) or monitoring efforts (Cole et al., 2024). Domain knowledge is implicit to policy, financing mechanisms and decision making and therefore acted upon. Further uncertainties, similar to model uncertainties in research, can originate from these informational representations if they create blind spots which prevent certain ecological or societal aspects from being managed (Goldstein et al., 2019) or cause these being managed inadequately. Therefore, it is crucial to address potential tradeoffs or adverse effects that may arise from simplified or differing perspectives and objectives reflected in domain representations of nature-based credits (Strange et al., 2024).

**A natural capital accounting approach: Biodiversity and ecosystem insurance value**

Strengthening environmental safeguards and leveraging ecological components for mitigation of physical risks, adaptation and resilience may become feasible by complementing carbon finance with more elaborate methods from natural capital accounting. These can account for some of the relations between ecosystem elements as demonstrated through scientific evidence and enable their integration on the level of accounting and financial decision-making.

A natural capital concept which satisfies conditions for integrating ecosystem components into natural capital accounting for physical risk mitigation and resilience efforts specifically, is the concept of biodiversity or ecosystem insurance value. Biodiversity or ecosystem insurance assigns a value specifically to the risk-reducing capacities of natural capital regarding physical risks such that they can be considered in risk-return assessments, adaptation and resilience co-benefits of projects, the pricing of carbon insurance or as complementary biodiversity credits (Baumgärtner 2007; Dallimer et al., 2020; Hahn et al., 2021; Quaas et al., 2019; Pascual et al., 2010). Therefore, these concepts can be particularly useful to leverage ecological components for physical risk mitigation and resilience, while enabling their consideration on the level of natural capital accounting. There exist varying approaches in the academic literature to conceptualizing biodiversity or ecosystem insurance value and resilience value for natural capital accounting. Although two broad directions have been distinguished - an economic conception and an ecological conception - a closer look reveals further differences.

**Economic conceptions of risk mitigation and adaptation capacities of natural capital**

In the economic conceptions of biodiversity or natural capital insurance as developed by Baumgärtner (2007), Baumgärtner and Strunz (2009) and Quaas et al. (2019), insurance is modeled similar to an investment into financial insurance. The insurance value of biodiversity or natural capital is the value derived by a risk-averse ecosystem user or investor from a marginal increase in biodiversity or ecological components to reduce the uncertainty of ecosystem output. This value is conceptualized through the reduction of the risk premium which the user is willing to pay to avoid uncertainty of ecosystem service provision. Biodiversity insurance is here conceptualized as inherently relational. It derives its insurance value from coupled ecological components such as biodiversity and ecosystem output in the form of provision of goods or services. While there exist other ways in natural capital accounting to relate use values and non-use values of ecosystems (Daily 1997; Daily et al., 1997; Kumar, 2010), the insurance value links risk-reducing ecological components to ecosystem output (Baumgärtner, 2007; Baumgärtner and Strunz, 2009; Quaas et al., 2019) and is hence additional to the ecosystem's output value. The internally stabilizing relations between different ecosystem processes and components have long been recognized (Folke et al., 2021; Folke, 2006; Holling, 1973; Isbell et al., 2015). The concepts of biodiversity insurance or natural capital insurance make use of these coupled components in the context of integrating and leveraging these for ecosystem management. Although biodiversity insurance is conceptualized as an ecosystem-internal risk-reducing mechanism, where relations between specific components of an ecosystem stabilize its provision of services and goods, this can be linked to the mitigation of ecosystem-external risks, such as securing the supply of food or clean water. Natural insurance in the economic sense has been modeled for agrobiodiversity (Quaas & Baumgärtner, 2010), quantified in case studies in agroecology (Di Falco and Chavas, 2006) and for non-timber products from forests (Pattanayak and Sills, 2001).

Another economic conception of risk mitigation through ecosystem-based adaptation and its valuation exists in the literature on disaster risk-reduction, climate resilience and adaptation (e.g. Marchal, 2019; UNDRR, 2020). Nature-based solutions are here evaluated regarding their capacity to reduce physical risk and reduce potential negative impact from climate- or nature-related risks and external hazards through their regulating ecosystem services (Daigneault et al., 2016; Narayan et al., 2017; Schwarze et al., 2011). Value is derived from the use of regulating ecosystem services such as water and climate regulation, natural hazard protection, disease and pest regulation. It can be assigned based on the damage costs that the provision of an ecosystem service such as flood mitigation through wetlands helps avoid and can be considered a part of an ecosystem's use value (Kumar, 2010). However, Dallimer et al. (2020) discuss multiple methods of valuation for different mitigation and adaptation measures, including risk reduction, prevention, avoidance, and adaptation measures. The two economic approaches to risk mitigation and adaptation through use of natural capacities can be divided into mechanisms that refer to internal capacities to reduce uncertainty of ecosystem output on the one hand, and regulating ecosystem services that buffer against external disturbances, on the other.

Two additional natural capital valuation approaches can be taken into account, especially when it comes to considering the long-term stability of ecosystems and sustainable management that is required to maintain their ability of ecosystem service provision. Both relate to valuing biodiversity under changing conditions and over time to ensure sustainable and stable ecosystem maintenance. Both therefore contribute to what can be conceptualized as an overall resilience value of ecosystems (Hahn et al., 2023). One is the real options approach which highlights how species substitutability can add value rather than diminish it (e.g. Kassar and Lasserre, 2004). Species act as assets that an ecosystem manager or society preserves for their potential to provide essential goods or services in the future or under changing conditions. The value derived from biodiversity lies in the option to substitute one species with another in response to changing conditions, which underlines the importance of preserving species that seem redundant. A related framework which specifically seeks to assign value to different ecological components as diverse assets of an ecosystem is the application of portfolio theory to ecosystems and biodiversity (Admiraal et al., 2013). This approach helps value biodiversity beyond its direct economic benefits as a stabilizing component for ecosystem service provision. Both approaches can help value not only the capacities to absorb and buffer against disturbances but also assign value to the underlying "resilience stock" of an ecosystem such as its absorptive and adaptive capacities and the ability to recover from disturbances and extreme events which are a necessary prerequisite to continue acting as a buffer.

Unarguably, ecosystem management practices contribute to either stabilizing or diminishing its capacities to maintain functioning in the face of disturbances and under changing conditions as well as to recover from disturbances. Academic literature has therefore argued for including them as adding value to risk mitigation and management (Daigneault et al., 2010; Herbert et al., 2022; Hurteau et al., 2008). If the concept of natural capital insurance is supposed to be extended into a full resilience value of ecosystems, as recently suggested (Hahn et al., 2023), selected social-ecological dimensions of its management and stewardship practices that specifically provide value regarding ecosystem-based risk mitigation and adaptation have to be integrated as adding value to it as well.

**Ecological insurance concept**

In contrast to the economic valuation of ecosystem-based risk reduction, the ecological conceptualization of ecosystem insurance (e.g. Yachi and Loreau, 1999) studies the many relations between different ecosystem components and their moderating, enhancing or stabilizing effects on one another, without assigning a monetary value to these processes or outcomes. Therefore, the ways in which a certain ecosystem component, such as biodiversity, stabilizes the provision of an ecosystem service, is often conceptualized in a multi-dimensional way rather than only by one measure of risk reduction such as reducing the variance of ecosystem output.

**Integration of approaches into resilience value as a natural capital accounting framework**

Rather than subsuming the different approaches under the umbrella term "insurance value", they can be categorized by their value assignment to various risk mitigation and adaptation capacities of natural capital. While Dallimer et al. (2020) provide an overview of natural capital risk mitigation mechanisms and valuation methods applied in empirical studies, the overview in *Figure 1* broadly distinguishes natural capital components and roles to which value is assigned. Eventually, they can be seen as mutually supportive and complementary.

1. Ecological studies of environmental insurance and risk mitigation effects provide the evidence base for natural capital accounting approaches that make use of these mechanisms when valuing biodiversity or natural capital in their specific risk mitigation capacity.

2. The economic approaches to nature-based risk mitigation refer to different fractions of natural capital:

    A. Biodiversity or natural capital insurance value as outlined by Baumgärtner (2007) and Quaas et al. (2019) leverages internal risk-reduction in ecosystems to safeguard ecosystem service and good provision and is valued according to the reduction of the risk premium that a marginal increase in natural capital stock provides.

    B. Ecosystem-based adaptation and disaster risk reduction (e.g. Marchal et al., 2019; Narayan et al., 2017; Schwarze et al., 2011) values provision of regulating ecosystem services for use of natural hazard protection or flood mitigation relating to ecosystem-external disturbances.

    C. Maintaining options and the application of portfolio theory help value biodiversity as underpinning the internally as well as externally risk mitigation mechanisms and establishing their maintenance over time (e.g. Admiraal et al., 2013; Kassar and Lasserre, 2004). Thereby, they help capture the resilience stock of an ecosystem underpinning the absorptive and adaptive capacities as well as the ability to recover from disturbances.

    D. Including practices in ecosystem management that specifically reduce risk from disturbances and increase adaptive capacity as well as capacities for recovery contributes to determining resilience value.

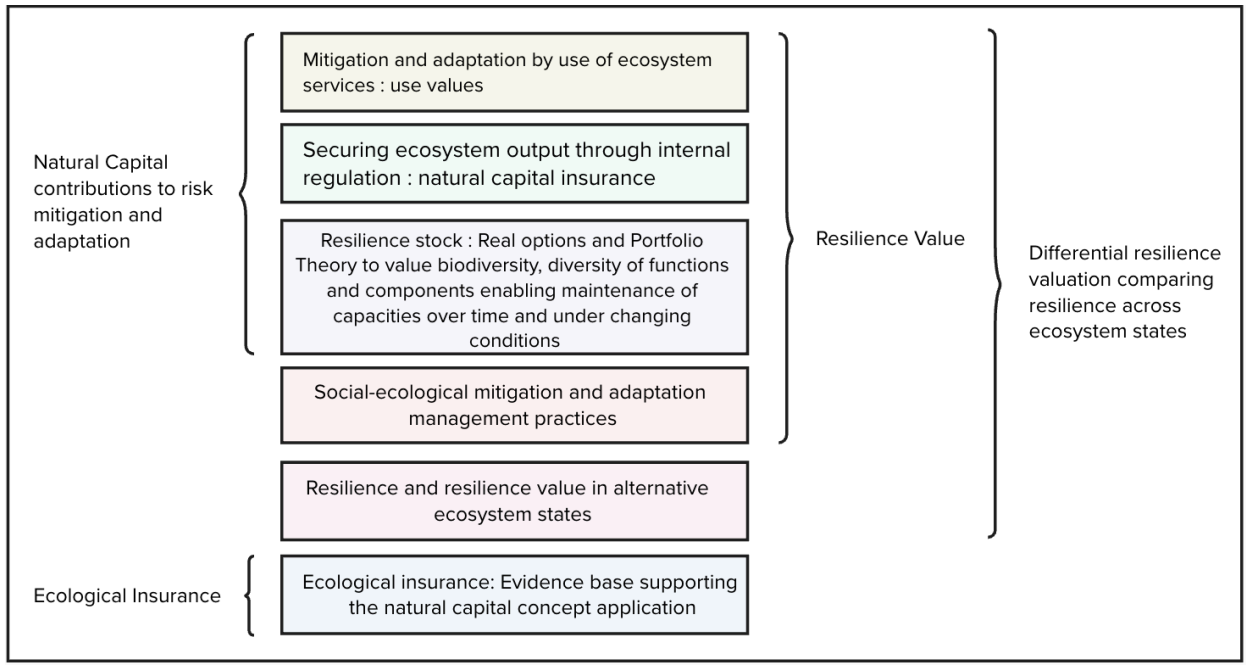

*Figure 1: Overview of natural capital components for ecosystem-based risk mitigation, adaptation and resilience and the role of ecological insurance as evidence base. Resilience value comprises social-ecological practices and natural capital. It can further be compared between alternative ecosystem states.*

A concept which can facilitate integration of different approaches is the insurance value as a full resilience value of an ecosystem as recently suggested by Hahn et al. (2023). A full resilience value is assumed to go beyond specific risk-reducing functions and establish a "general resilience value" of an ecosystem in the face of multiple disturbances and for multiple stakeholders over time. The authors also suggest addressing the increasing ecosystem vulnerabilities which happen through slow erosion of the resilience stock. When it comes to resilience value, the possibility arises that an ecosystem reaches a threshold and switches states or regimes. Another valuation perspective therefore emerges from the comparison of the resilient functioning of an ecosystem across different states or regimes (Pascual et al., 2010; Walker et al., 2009a, 2009b, Walker and Meyers, 2004). A full resilience value can serve as a context-sensitive framework in natural capital accounting that aggregates several value components that a social-ecological system contains specifically for resilience and adaptation relative to its current state and under certain asset- and project-specific risk exposure conditions.

**Application of biodiversity insurance value and resilience value to carbon project development**
All conceptions of nature-based risk mitigation and insurance values can in theory be applied to a carbon project from forestry. In practice, such an effort poses challenges in operationalization of the natural capital concepts as well as for data acquisition, modeling and simulation and data analysis required to assess the specific risk mitigation potential of ecosystems and management practices relative to risk exposure.

However, following the conclusion from the previous section that economic concepts of nature-based risk mitigation require evidence of ecological insurance effects, Baumgärtner's model (2007) is particularly well suited for capturing the stabilizing function of tree species diversity for carbon storage and sequestration (e.g. Mori et al., 2021; Silva Pedro et al., 2015; Weiskopf et al., 2024). Baumgärtner's description specifically focuses on the role of biodiversity as insurance for reducing uncertainty of a specific ecosystem service output. Additionally, the case of carbon credits from forestry aligns with further application conditions for his model (Baumgärtner, 2007, p.110):

1. Ecosystem users and investors are risk-averse.
2. Financial insurance for carbon projects is available.
3. The level of biodiversity impacts the probability distribution of the ecosystem service and, thus, of income from carbon credits, and
4. The insured and carbon insurer have the same *ex ante* knowledge about the distribution of carbon sequestration provision.

Unarguably, the exact insurance effect of biodiversity within a forest ecosystem has to be specified relative to the particular geography, type, use, age, management and disturbance history, composition and structure of the forest ecosystem and its specific social-environmental conditions. Moreover, the specific metrics such as α-diversity, species richness, abundance, and the like and statistical measures in which relations are measured, have to be adjusted accordingly. Further, if biodiversity as an insurance is specifically valued in its role to secure carbon sequestration, safeguarding measures are required to ensure that this does not lead to tradeoffs in forest management to prioritize stewardship for the stability of the carbon cycle at the expense of biodiversity or other benefits. The insurance value of biodiversity only values a specific role of biodiversity in an ecosystem and does not account for its entire value. Studies exist which outline forest ecosystem conditions in favor of both objectives, carbon storage and biodiversity, relative to a specific geographic context which can be supportive for natural capital concepts (Springer et al., 2024).

In the remaining sections, Baumgärtner's model of biodiversity insurance and its potential application to the carbon project case is discussed. This is followed by a brief description of how a full resilience value as outlined by Hahn et al. (2023) goes beyond the internal mitigation concept of an ecosystem insurance. The concluding sections provide an outlook on the necessities for improved modeling and operationalization of multiple approaches to enable their practical applicability.

**Application of biodiversity insurance value to the biodiversity-carbon relationship in carbon forest management**

Baumgärtner (2007) conceptualizes the insurance value of biodiversity as the benefit derived by a risk-averse ecosystem manager from reducing the risk premium R associated with the manager's uncertain income from ecosystem service provision. In Baumgärtner's model, the manager derives their income y from ecosystem service provision s e.g. carbon sequestration. This ecosystem service provision is modeled as a normally distributed random variable with mean $\mu s$ and standard deviation $\sigma s$ and so is the manager's income. It is assumed that the mean level of ecosystem service provision increases and its standard deviation decreases in proportion to the level of biodiversity v in the ecosystem. Hence, investing in biodiversity insurance reduces the variance of ecosystem service outcomes. As mentioned, the relation of biodiversity and carbon within a specific forest ecosystem can be further specified with respect to a case study. As an orientation to facilitate applicability, it is here assumed that tree species diversity increases and stabilizes carbon storage in a forest ecosystem (Liang et al., 2016; Mori et al., 2021; Qiao et al., 2023; Silva Pedro et al., 2015; Weiskopf et al., 2024; ).

The distribution of the ecosystem service is hence conditional on the level of biodiversity v in the ecosystem. Investing in biodiversity comes at a cost C. The net income y of the ecosystem manager when using biodiversity as a natural insurance is the benefits B they derive from the ecosystem service s minus the costs C of their investment into biodiversity v:

$$y = B(s) - C(v) = s - C(v).$$

**Risk aversion and conditional variability**

Baumgärtner's concept of biodiversity insurance value V depends on (1) the conditional variability of the ecosystem service, here carbon sequestration, on biodiversity v and (2) the degree of risk aversion of the ecosystem manager. Investors and project developers in carbon finance can be assumed to be risk averse which is why financial insurance for carbon projects exists. Baumgärtner captures these assumptions in his Proposition 1:

$$V(v) = -\rho \sigma s(v) \sigma s'(v) > 0.$$

where $\rho$ is the ecosystems manager's degree of risk aversion, $\sigma s(v)$ is the standard deviation at a level of biodiversity v and $\sigma s'(v)$ the derivative of the standard deviation which indicates the rate of change of the standard deviation $\sigma s$ with respect to changes in biodiversity level v. The insurance value V of biodiversity v increases with the degree of the ecosystem manager's risk aversion and with the degree of sensitivity of the standard deviation of ecosystem services on changes in biodiversity. If this sensitivity measure is high, the power of biodiversity to regulate uncertainty of ecosystem service provision increases, and hence its insurance value increases.

**Uncertainty definition**

The uncertainty of an ecosystem is modeled by Baumgärtner as an infinite set of lotteries. The risk premium R represents the amount a decision maker is willing to pay to avoid the uncertainty of a particular lottery. The higher the risk aversion of the manager, the higher the amount they are willing to pay to avoid risky outcomes. In Baumgärtner's approach, the risk premium R of the lottery depends on the level of biodiversity v in an ecosystem. More specifically, given the conditionality of ecosystem service provision on the level of biodiversity, choosing a particular level of biodiversity, the ecosystem manager determines a distribution of an ecosystem service, e.g. carbon sequestration, which then determines the distribution of their income, e.g. income from carbon credits based on the ecosystem's carbon uptake.

The insurance value V of biodiversity v is then defined by Baumgärtner as the change of the risk premium R due to a marginal change in the level of biodiversity v:

$$V(v) := -R'(v).$$

By changing the biodiversity level, the manager can potentially reduce uncertainty - change the lottery for the better - and hence reduce the amount of R. The insurance value derives from the reduction in R that an investment into biodiversity provides. Applied to the case of land and forest management activities for carbon, depending on the level of biodiversity available and depending on the available management options, investors or project managers face the choice of investing more in biodiversity to reduce the variance of carbon outcomes.

**Optimization with and without financial insurance**

In Baumgärtner's model, the ecosystem manager seeks to maximize his expected utility by choosing an optimal level of biodiversity. Further, they seek to allocate resources efficiently to maximize their expected utility considering the benefits and costs associated with biodiversity. Baumgärtner first examines how a risk-averse ecosystem manager makes use of the biodiversity insurance function if no financial insurance is available. In this situation, the manager keeps investing in biodiversity up to the point where the marginal benefit from an increase in biodiversity in terms of ecosystem service provision and biodiversity insurance value equals the marginal cost of biodiversity maintenance or enhancement. This characterizes the optimal level beyond which any additional investment yields a diminishing or zero net gain.

When financial insurance is an option, the ecosystem manager decides on both the level of biodiversity and the extent of financial insurance coverage to maximize their expected utility. In Baumgärtner's model, financial and biodiversity insurance act as substitutes and can both help reduce the overall risk or variability of returns from natural resources. The ecosystem manager's net income is then defined by several factors: the provision of ecosystem services, the costs and benefits from biodiversity insurance, and the expenses for financial insurance coverage. The manager allocates resources optimally between investing in biodiversity insurance and in financial insurance, while the investment into financial insurance is dependent on its real costs. Baumgärtner eventually concludes that a risk-averse ecosystem manager will always opt for a lower level of biodiversity when financial insurance is available, compared to a scenario without financial insurance. This might be due to the perceived higher reliability and the more quantifiable way to manage risk of financial insurance. It however suggests that without additional incentives or regulations, investors may overlook the ecological benefits of biodiversity in favor of financial instruments, which stresses the importance of policies that promote or incentivize biodiversity investment. For the case of nature-based carbon projects, this implies that the access to financial carbon insurance could disincentivize an investment into biodiversity as a natural insurance mechanism if the insurance does not consider risk mitigation measures using natural resources or if there is no policy in place that incentivizes investment into natural capital as an insurance mechanism. Such cases have been discussed in financial insurance of natural resources, e.g. in farming (RFSI, 2024), where financially insuring crops might disincentivize sustainable farming. This suggests that natural insurance should be considered and valued when financially insuring natural resources and their management. The substitutability assumption in Baumgärtner's model serves to highlight the potential tradeoffs from availability of financial insurance and natural capital insurance. However, the assumption is certainly questionable due to the limited substitutability of financial and natural capital, including the long-term effects of these investment choices.

**Including biodiversity insurance in natural capital accounting for carbon insurance or co-benefits**

One way to leverage biodiversity insurance for carbon projects within natural capital accounting is to adjust the risk premium for financial carbon insurance—where not already done—based on investment in biodiversity enhancements, such as higher tree species diversity for biodiversity insurance. Another option is to include a biodiversity or broader ecosystem insurance value component within the carbon credit itself as an additional co-benefit, without necessarily replacing a standalone biodiversity component. Alternatively, nature-based carbon could be complemented with biodiversity insurance credits.

Following Baumgärtner's model, the description here specifically relates carbon sequestration and tree species diversity as the latter providing insurance for the former which can be operationalized and applied in more detail to a real carbon project from forestry. Clearly, there are other ecosystem services besides carbon sequestration or storage - such as protecting against diseases, enhancing belowground biodiversity, stabilizing water storage - that an increase in biodiversity in a forest potentially enhances and stabilizes. Secondly, there might exist other forms of risk reduction than reducing the variance of carbon sequestration provision which enhancement of forest diversity provides. Thirdly, there are other ecological components that have to be leveraged to provide broader project resilience in the face of a diverse range and different types of climate-related, nature-related and social risks. A broader resilience value takes other ecosystem components and their conditionalities and causal links into account.

**Developing a full resilience value : accounting for subjective risk preferences and objective risks**
Hahn et al. (2023) question the economic conception of biodiversity or natural capital insurance as outlined by Baumgärtner (2007) or Quaas et al. (2019) under the assumption that, given that climate- and nature-related risks are what they call objective risks, the subjective risk preferences become obsolete or just a fraction of these risks. Consequently, in the face of such objective risks, the insurance value of ecosystems is, regardless of risk preferences, the "full resilience value", meaning it also addresses the objective risks of climate change, nature-related risks and social risks. With their distinction between "subjective" and "objective" risks, Hahn et al. (2023) seem to refer to the varying existence conditions of different types of risks. Objective risks may be conceptualized as risks that exist independently of their perception as risks from certain agents, whereas subjective risks exist dependent on the specific conditions and choices of a certain risk-taking perspective. An agent might ignore or deny the existence of risks of climate change and yet these risks may materialize when their assets are hit by a climate-change induced flood. Acknowledging these risks is a choice, but being confronted with them isn't completely in the control of the agent. However, the choice to help mitigate or reduce these risks through ecosystems and nature-based solutions is available as a management option. In a scenario where an agent faces subjective risks, in contrast, an agent can fully decide whether they want to take on a certain risk, e.g. by making an investment, or not.
A further way of understanding the distinction between objective and subjective risks is regarding whether the risk perspective only considers private costs and benefits or also social costs and benefits. Under a subjective risk perspective, only the private costs and benefits of risks are considered, whereas under an objective risk perspective, social costs and benefits are included. In any case, the authors leverage the public good nature of ecosystems and its implication of being non-excludable. Therefore, the risk reducing services of ecosystems such as drought or flood mitigation are non-excludable, too, and do not only mitigate subjectively chosen risks. They are more inclusive by nature, mitigating risks of different types for all stakeholders. This is another way of understanding the distinction of subjective and objective risks, relative to whether the risk-reducing services of nature are conditional on investment and management choices of an ecosystem manager (subjective risks) or not (objective risks), because they are just a part of how ecosystems function.
Therefore, Hahn et al. (2023) argue for assessing the full resilience value of ecosystems which aims to be more comprehensive, including the mitigation of subjective risks through increasing the natural capital stock and of objective risks by delivering the risk mitigation through ecosystem services that are non-excludable from a public goods perspective. Further, as the authors point out, the full resilience value includes a resilience stock which ensures provision of ecosystem goods and services over time and under changing conditions. A full resilience value subsequently not only includes mitigation of and adaptation to different types of risks, but it also assigns value to more than one natural capital component and different processes required to fulfill these roles.

**Challenges in operationalization of resilience value**

To meet the requirements of mitigating internal and external risks, as well as subjective and objective risks, the full resilience value of natural insurance as outlined by Hahn et al. (2023) needs to comprise an aggregate of risk-reducing services and components relative to multiple risks, regardless of a specific user perspective. There exist challenges regarding the operationalization and applicability of a full resilience value, but efforts to do so can build on the various natural capital approaches to valuing nature-based physical risk mitigation and adaptation that are already available: the valuation of nature-based solutions in disaster risk reduction, the internal natural capital insurance value, as well as the portfolio and the real options approach to valuing the maintenance of a "resilience stock" over time and under changing conditions.

The inclusion of nature- and climate-related risks into conceptions of nature-based risk mitigation additionally necessitates a change in their statistical operationalization. More specifically, buffer or insurance effects can stabilize ecosystem services provision not only relative to normally distributed ecosystem services, such as in the model described by Baumgärtner (2007), but also relative to their distributions under and after extreme climate- and nature-related events and potential regime shifts. The operationalization of resilience value of an ecosystem undergoing extreme events not only requires a dynamic and variable conceptualization as e.g. in Quaas et al. (2019) but also an improved understanding of how risks are compounded by environmental factors or systemic vulnerabilities. It further requires understanding the ecosystem responses to climate events (Anderson-Texeira et al., 2022) and the interactions of ecosystems with climate conditions such as the coupling of the carbon and water cycle (Gentine et al., 2019). Further, it requires complementary assessment of ecosystem vulnerabilities and risk exposure.

**Resilience value in carbon projects from forest-related activities**

Increasing resilience value within a carbon project implies implementation of adaptation and resilience measures that not only mitigate risks of uncertain ecosystem service provision, such as by enhancing tree species diversity, but also include strengthening the overall resilience of ecosystems in the face of disturbances. Although this results in enhancing resilience for multiple stakeholders, it includes stabilizing ecosystem provision for the risk-averse ecosystem investor or manager. A customized approach to natural capital valuation for resilience specifically in carbon forest management could integrate resilience and adaptation practices across potential disturbances and management objectives that are considered relevant relative to a specific project context and the risk exposure of that project. Suggestions in the academic literature have been made to include mechanical thinning or fuel reduction treatments for reducing risks from forest fires (Herbert et al., 2022), to consider sustainable water use and storage to mitigate potential impacts from drought and consider coupling of the carbon and the water cycle (Martinez-Sanchez et al., 2022;) as well as biodiversity enhancement to secure carbon storage and sequestration (Springer et al., 2024; Weiskopf et al., 2024). These environmental measures can then be complemented by social resilience considerations supporting local communities and the sustainability of their livelihoods. Existing frameworks such as the World Bank Resilience Rating system (2021) can further help design and evaluate project-specific strategies to increase carbon project resilience in the face of climate change which can eventually be considered during risk and risk mitigation assessment, factored into project co-benefits, and into carbon insurance pricing or potentially issued as additional biodiversity credits.

**Conclusion**

Climate adaptation and broader resilience management is an important topic in risk assessments and risk mitigation efforts in the financial, public and corporate sector already (Baer et al., 2023b; Mullan, 2024; Mullan and Ranger, 2022; Spacey Martin et al., 2024; World Bank, 2024). Addressing risks in their double materiality, comprising financial, social and physical risks, requires integrating different means into comprehensive risk mitigation strategies. Those strategies can leverage and incorporate natural insurance through ecosystem based risk-mitigation similar to nature-based solutions that are used in ecosystem-based disaster risk reduction (UNDRR, 2020). In order to make the use of natural insurance financially viable, concepts in natural accounting have to be applied that can assign a value to the different ways in which ecosystems contribute to reducing, mitigating and adapting to risks - through natural capital insurance, a resilience stock and ecosystem service provision.

This also applies to the design and implementation of nature-based carbon projects. Research indicates that nature-based carbon projects require specification of social and environmental safeguards to integrate natural capacities and management activities for physical risk mitigation of climate change and nature-related risks (Wissner & Schneider, 2022). Although the concept of environmental and social safeguards is in use, there are shortcomings in its implementation (Haya et al. 2024). And even if nature-based risk mitigation is implemented on the project level, the question arises whether and where it is valued in finance and natural capital accounting that can account for nature's contributions specifically to risk mitigation and insurance pricing.

The concepts of biodiversity or ecosystem insurance and ecosystem resilience value (Baumgärtner, 2007; Hahn et al., 2023; Primmer and Paavola, 2021; Quaas et al., 2019) provide valuable starting points to integrate environmental components into natural capital accounting for insurance pricing or physical risk mitigation. The economic approach to ecosystem-internal natural capital insurance (Baumgärtner, 2007; Quaas et al., 2019) to mitigate uncertainties of specific ecosystem service provision can be valuable in investment decision-making contexts as well as to integrate nature insurance value into financial insurances because it is modeled similar to a financial insurance mechanism. The economic approach modeled on ecosystem provision through nature-based solutions, in contrast, allows to assign value to risk reduction through regulating ecosystem services by considering avoided damage costs. Valuing a resilience stock that ensures provision of risk mitigation capacities over time and under changing conditions requires a portfolio or real options approach. Studies of ecological insurance effects provide an evidence base for applying natural capital concepts based on scientific results. The resilience value highlights the necessity to distinguish different types of risks and the varying requirements for their reduction and mitigation, but is in need of further methodological improvement regarding its operationalization and potential applicability.

Acknowledgments: I thank Sven von Vitorelli for providing useful comments to an earlier version of this text.

Funding: None.

Competing interests: The author and Leipzig University have no commercial interest in carbon credits.

Data and materials availability: The manuscript contains no data.